\documentclass[aps,pra,floatfix,twocolumn,nofootinbib,showpacs]{revtex4}
\usepackage{amsmath}
\usepackage{amssymb}
\usepackage{graphicx}
\usepackage{dcolumn}
\usepackage[sort&compress]{natbib}
\usepackage{bm}
\usepackage[dvips]{epsfig}
\newcommand{\BEA}{\begin{eqnarray}}
\newcommand{\EEA}{\end{eqnarray}}
\renewcommand{\d}{{\rm d}}
\newcommand{\comment}[1]{}
\newcommand{\bx}{{\bf x}}
\newcommand{\bp}{{\bf p}}
\newcommand{\bq}{{\bf q}}
\newcommand{\mless}{\,\prec_{\bp}\,}
\newcommand{\mgrt}{\,\succ_{\bp}\,}
\newcommand{\x}{x^{\downarrow}}
\newcommand{\p}{p^{\downarrow}}
\newcommand{\wa}{\widetilde{a}}

\newcommand{\G}{\Gamma}
\begin{document}

\title{Le Chatelier principle in replicator dynamics}
\author{Armen E. Allahverdyan$^{1)}$ and Aram Galstyan$^{2)}$}

\address{$^{1)}$Yerevan Physics Institute, Alikhanian Brothers Street 2, Yerevan 375036, Armenia \\
$^{2)}$  USC Information Sciences Institute, 4676 Admiralty Way, 
Marina del Rey, CA 90292, USA}

\begin{abstract} The Le Chatelier principle states that physical
equilibria are not only stable, but they also resist external
perturbations via short-time negative-feedback mechanisms: a
perturbation induces processes tending to diminish its results. The
principle has deep roots, e.g., in thermodynamics it is closely related
to the second law and the positivity of the entropy production. Here we
study the applicability of the Le Chatelier principle to evolutionary
game theory, i.e., to perturbations of a Nash equilibrium within the
replicator dynamics. We show that the principle can be
reformulated as a majorization relation. This defines a stability notion
that generalizes the concept of evolutionary stability. We determine
criteria for a Nash equilibrium to satisfy the Le Chatelier principle
and relate them to mutualistic interactions (game-theoretical
anticoordination) showing in which sense mutualistic replicators can be
more stable than (say) competing ones. There are globally stable Nash
equilibria, where the Le Chatelier principle is violated even locally:
in contrast to the thermodynamic equilibrium a Nash equilibrium can
amplify small perturbations, though both this type of equilibria satisfy
the detailed balance condition.

\end{abstract}

\pacs{87.23.-n, 02.50.Le, 87.23.Cc, 87.23.Kg }






\maketitle

\section{Introduction}

\comment{Lenz's law \cite{edu}: {\it An induced current is always in such a direction as to
oppose the motion or change causing it}.}

{\it An external influence disturbing an equilibrium state of a system
induces processes tending to diminish the results of the disturbance.}
This is the qualitative content of the principle formulated by Le
Chatelier for chemical reactions \cite{edu}. Earlier statements of the
principle are reviewed in \cite{phi}.  Braun related the principle to
the second law and extended it to general thermodynamic equilibria
\cite{landau,gilmore}. In contrast to the second law, the Le Chatelier
principle is more operational and more intuitive; hence it became the
main tool for predicting the response of physical and chemical
equilibrium \cite{edu,landau,gilmore}. Applications of the principle go
well beyond physics and chemistry: in economics it is used for the
analysis of price equilibrium \cite{econo}; in ecology for explaining
qualitatively the ecosystems growth \cite{nature,jorge}, and also for
quantifying the threshold of allowed influences of civilization on the
environment \cite{gorshkov}. The principle gave rise to the concept of
{\it homeostasis}, a state of an organism or construction that is stable
due to compensatory mechanisms \cite{phi}. 

The issue of stability is also central to population dynamics and
ecology: to a large extent these disciplines emerged and developed
around various aspects of the stability notion and its relations to
diversity, productivity, complexity {\it etc} \cite{svi}; see
\cite{ives} for a recent review. The question of understanding pertinent
forms of stability is presently more pending than ever in face of various
environmental issues and global changes. 

Our purpose here is to develop a formalization of the Le Chatelier
principle for a game-theoretical population dynamics and to show that it
leads to a notion of stability that allows to make new predictions. In
particular, the Le Chatelier principle differs from the asymptotic
stability \cite{arnold}, where|once a (small) perturbation is over|the
equilibrium is recovered after a sufficiently large time (relaxation)
\footnote{But asymptotic stability is more than relaxation. It includes
the notion of Lyapunov stability: for {\it any} vicinity $A$ of the
fixed point, there is its subset $B$ such that the motion that starts in
$B$ never leaves $A$. There are examples of relaxing, but unstable fixed
points \cite{arnold}.  }. It is however not excluded that in the course
of relaxation the perturbation will be transiently amplified, a scenario
forbidden by the Le Chatelier principle. While the asymptotic stability
indicates on long-time negative feedback (perturbation does decay sooner
or later), the principle means a {\it short-time} negative feedback. 

Our setup is the the replicator dynamics, a basic description of the
Evolutionary Game Theory (EGT) and population dynamics
\cite{zeeman,hofbauer,sandholm,svi}. This dynamics originated in 1930's
as a population approach to genetics \cite{hofbauer}. Later on it merged
with game theory \cite{sandholm} and with mathematical ecology
\cite{svi}. It has a wide range of
applications from biology, ecology and economics
\cite{hofbauer,sandholm,svi} to physics \cite{armen_mahler} and
combinatorial optimization \cite{menon}. EGT describes large
populations of agents (humans, animals, microbes, genes) interacting via
playing a game. Changes in these populations are driven either by the
application of decision rules by individual agents, or by natural
selection via different reproduction rates
\cite{zeeman,hofbauer,sandholm}. EGT related the Nash equilibrium of a
static game to attractors of dynamics \cite{zeeman,hofbauer,sandholm},
thereby likening the Nash equilibrium to the thermodynamic equilibrium.
EGT also put forward the concept of evolutionary stability, a widely
used notion related to stability with respect to invasions
\cite{zeeman,hofbauer,sandholm}. 

Here we show that {\it i)} for perturbations of an asymptotically stable
Nash equilibrium the Le Chatelier principle can be reformulated as a
majorization relation.  {\it ii)} the criterion for the principle
relates to mutualistic [cooperative] interactions between agents
(game-theoretical anticoordination). {\it iii)} A weaker formulation of
the Le Chatelier principle, which demands that a perturbation is at
least not amplified \footnote{More precisely, a violation of the weak Le
Chatelier principle means that the perturbation is amplified, and
simultaneously the perturbed system is destablized; see below. },
provides a stability concept with a wider applicability than the
evolutionary stability, e.g., the principle can be applied to permanent
replicators that do not possess an asymptotically stable state. {\it
iv)} Non-mutualistic (competitive or prey-predator) interactions can
violate even the weak formulation of the principle, i.e., small
perturbations over an asymptotically stable Nash equilibrium can be
amplified. This is impossible for thermodynamic equilibria and shows
that Nash equilibria can be endowed with positive feedback. 

\comment{
To put our questions in a wider perspective, note that the idea of the
Le Chatelier principle, with the equilibrium tending to recover in short
times, relates to the notion of {\it robustness} widely discussed in
system biology \cite{kitano,lesne,jen}. Robustness broadens the notion
of stability in several respects, e.g., generalizing it to multiple
perturbations or accounting for both state and structural perturbations
\cite{jen}. These two aspects are harbored by the Le Chatelier principle
showing that the robustness of population dynamics can be understood via
the principle.}

This paper is structured as follows: Section \ref{sl} introduces the Le
Chatelier principle and explains its meaning: \ref{weak} states a weaker
version of the principle and shows its relations with multiple
perturbations. \comment{ Section \ref{struc} outlines connections
between the principle and structural perturbations.} Section \ref{repli}
recalls the replicator dynamics and the main features of the Nash
equilibrium. In particular, section \ref{dbala} explain why the notion
of Nash equilibrium is indeed similar to the thermodynamic equilibrium,
while section \ref{ESS} reviews the concept of evolutionary stability
and relates it with the notion of mutualism as the whole. Section
\ref{crit} obtains the criteria of the Le Chatelier principle for the
replicator dynamics. Section \ref{n3} studies the principle for concrete
population games. The last section concludes and makes connections with
previous works and ideas. 

\section{Le Chatelier principle}
\label{sl}

\subsection{The formulation}
\label{gen}

\comment{
We already mentioned above that the Le Chatelier principle can be
derived from the entropy maximization aspect of the second law of
thermodynamics \cite{landau}. However, the way this derivation is done
in thermodynamics \cite{landau} and also economics \cite{econo}
conseals the fact that the Le Chatelier principle is a stronger
statement than the second law and that it can be formulated in a setup
that does not refer directly to the entropy maximization, and 
thus applies to open systems in a wider context. }

The standard formulation of the principle in equilibrium thermodynamics
refers to {\it small} perturbations of {\it extensive} variables
(volume, number of particles for a certain substance, {\it etc})
\cite{gilmore}. The linear non-equilibrium thermodynamics formulates the
principle via fluxes and forces again for small deviations from
equilibrium \cite{groot,trincher}. In contrast, we need a formulation
that works within the non-equilibrium statistical mechanics approach,
where the basic quantities are probabilities of states satisfying
certain equations of motions. Such a formulation should refer to not
necessarily small perturbations of the autonomous equations of motion. 

Let a system is described by a probability vector
from the simplex ${\bf S}_n$:
\BEA
\label{ko}
\bx\in{\bf S}_n ~:~
\bx^{\rm T}\equiv \{x_k\geq 0\}_{k=1}^n, \qquad {\sum}_{k=1}^n x_k =1,
\EEA
where the probabilities $x_k$ refer to occupation of various states in
the population, and where $\bx^{\rm T}$ means transposition of the column
$\bx$. In statistical physics the index $k$ typically refers to (free) energy
levels or to phase-space cells \cite{landau}. In population dynamics $k$
refers to groups of agents having certain common features (traits).
The interior of ${\bf S}_n$ is given by
$x_k>0$ and ${\sum}_{k=1}^n x_k =1$. 

Let $\bx(t)$ satisfy a continuous-time master equation:
\BEA
\label{rota}
\d x_i(t)/\d t\equiv\dot{x}_i = {\sum}_{k=1}^n W_{ik}[\bx]x_k, ~ {\sum}_{i=1}^n W_{ik}[\bx] =0,
\EEA
where $W_{i\not =k}[\bx]\geq 0$ is the transition probability $k\to i$,
and ${\sum}_{i=1}^n W_{ik}[\bx] =0$ follows from ${\sum}_{k=1}^n x_k =1$.
Eq.~(\ref{rota}) is the base for non-equilibrium statistical mechanics,
chemical kinetics and population dynamics \cite{kampen}. $W_{i\not
=k}[\bx]$ may depend on $\bx$ (non-linear case), as in the Boltzmann
equation \cite{kampen} or in the replicator equation
\cite{hofbauer,sandholm}. When $W_{i\not =k}$ does not depend on $\bx$
(linear case), (\ref{rota}) refers to the one-time probability
a continuous-time Markov process
\footnote{The case when $W_{ik}[\bx]$ does depend on $\bx$ is frequently described in literature
as a non-linear Markov process \cite{frank,kolokol}. There are pros and cons for this. If one stays at the level
of the one-time probability $\bx(t)$ the linear and non-linear situation share several common features \cite{frank,kolokol,franko}, e.g.
in both cases the future is determined by the present only (no influence of the past is to be taken into account), and in both cases there
is at least one stationary (i.e. time-independent) probability vector (a consequence of the Bauer's fixed point theorem).
However, it was stressed that a stochastic process is determined by its multi-time probabilities, and from this viewpoint
the non-linear equation (\ref{rota}) does not correspond to any well-defined stochastic process \cite{mccauley}.}. 

Let $\bp^{\rm T}=\{p_k>0\}_{k=1}^n$ be an asymptotically stable rest-point
of (\ref{rota}). A sudden perturbation at time $t$ brings the system
from $\bp$ to some $\bx(t)$ from the attraction basin of $\bp$. The
perturbation is quantified via ratios $\{x_k(t)/p_k\}_{k=1}^n$. We
re-number these ratios as
\BEA
\label{kia}
{\x_1(t)}/{\p_1}\,\,\geq\,\, {\x_2(t)}/{\p_2}\,\,\geq\,\,\ldots \,\,\geq\,\,{\x_n(t)}/{\p_n}.
\EEA
Thus $\x_1(t)$ [$\x_n(t)$] is the largest [smallest] component of perturbation. 
The solution of (\ref{rota}) is a smooth function of time. Hence the ordering
(\ref{kia}) holds at time $t+\epsilon$ with a small $\epsilon>0$:
\BEA
{\x_1(t+\epsilon)}/{\p_1}\,\geq\, {\x_2(t+\epsilon)}/{\p_2}\,\geq\,\ldots \,\geq\,{\x_n(t+\epsilon)}/{\p_n}.
\label{brox}
\EEA
If all inequalities in (\ref{kia}) are strict, the transition from
(\ref{kia}) to (\ref{brox}) is obvious, since $\epsilon\to 0$ and $\bx(t)$
is a continuous function. If some of them are equalities, ordering
(\ref{kia}) is not unique, but it can be chosen such that (\ref{brox})
holds\footnote{Note that although the ordering (\ref{kia}) is thereby
conserved for short times, it need not be conserved for long times,
simply because this ordering is not unique if some relations in
(\ref{kia}) are equalities. Hence it is possible (for $t_1<t_2<t_3$)
that $\bx(t_1)$ and $\bx(t_2)$ are ordered simularly (in the sense of
(\ref{kia})), $\bx(t_2)$ and $\bx(t_3)$ are also ordered simularly, but
$\bx(t_1)$ and $\bx(t_3)$ are not ordered simularly. }.  Perturbations
can have external origin, or be related to intrinsic noise. In the
latter scenario (\ref{rota}) arises as a deterministic approximation to
a random dynamics \cite{sandholm}, so that the remnants of noise can be
described as rare random perturbations. 

To formulate the Le Chatelier principle gradually assume first that
$n=3$. According to (\ref{kia}), $x_1$ is the strongest component of the
perturbation. Short-time negative feedback means that it decays:
$\x_1(t+\epsilon)\leq \x_1(t)$. Likewise, $\x_3(t)$ is the smallest
component of the perturbation. It may be close to zero making the
corresponding group vulnerable to extinction due to intrinsic noise
\cite{sandholm}. The Le Chatelier principle requires negative feedback
mechanisms increasing $\x_3(t)$: $\x_3(t+\epsilon)\geq \x_3(t)$.
Combining this with the above condition on $\x_1(t)$ we obtain
(\ref{sa}) for $n=3$. For $n>3$ we note that we may define the sum
$\x_1(t)+\x_2(t)$ as the largest component of the perturbation and
demand its decay in time.  Likewise, $\x_n(t)+\x_{n-1}(t)$ can be taken
as the smallest component demanding its non-decrease in time.  Hence we
come to the following formulation for the Le Chatelier principle to hold
for times $t\in [0, {\cal T}]$ for perturbation $\bx(t)$ ($\epsilon\to
0+$):
\BEA
\label{sa}
{\sum}_{k=1}^m \x_k(t+\epsilon)\leq {\sum}_{k=1}^m \x_k(t), \quad 1\leq m\leq n,
\EEA
with the equality for $m=n$.
Taking $\epsilon\to 0$ in (\ref{sa}) we get:
\BEA
\label{sar}
{\sum}_{k=1}^m \dot{x}^\downarrow_k(t)\leq 0, \quad 1\leq m\leq n.
\EEA
This formulation can be presented in an equivalent form.
Eqs.~(\ref{sa}, \ref{kia}, \ref{brox}) imply 
\footnote{\label{foot0}Define $A_k(t)\equiv {\sum}_{l=1}^k \x_l(t)$ and
$Q_k\equiv [f(\frac{\x_k(t)}{\p_k})-f(\frac{\x_k(t+\epsilon)}{\p_k})]/[\frac{\x_k(t)}{\p_k}-\frac{\x_k(t+\epsilon)}{\p_k}]$.
Note that $Q_{1}\geq ...\geq Q_n$ due to the convexity of $f(y)$ and to (\ref{kia}, \ref{brox}). Put $A_0(t)=A_0(t+\epsilon)=0$. Now
$\sum_{k=1}^n\p_k\left[f(\frac{\x_k(t)}{\p_k})-f(\frac{\x_k(t+\epsilon)}{\p_k})    \right]=\sum_{k=1}^n Q_k\{
A_k(t)-A_{k-1}(t)-[A_{k}(t+\epsilon) -A_{k-1}(t+\epsilon)]\}
={\sum}_{k=1}^{n-1}[A_k(t)-A_{k}(t+\epsilon)]\,[Q_k-Q_{k+1}]\geq 0$ due to (\ref{sa}) and $A_n(t)=A_n(t+\epsilon)$ \cite{olkin}. 
} that for any convex [$f''(y)\geq 0$] function $f(y)$ one
has \cite{ruch_mead,ruch_trio,olkin}:
\BEA
\label{san}
{\sum}_{k=1}^np_k f(x_k(t+\epsilon)/p_k) \leq {\sum}_{k=1}^np_k f(x_k(t)/p_k),
\EEA
which after taking the limit $\epsilon\to 0+$ reduces to 
\BEA
{\sum}_{k=1}^n\dot{x}_k f'(x_k(t)/p_k)\leq 0,\qquad f'(x)=\d f/\d x.
\label{ra}
\EEA
The converse holds as well \cite{ruch_mead,ruch_trio,olkin}: (\ref{san}, \ref{kia}, \ref{brox}) 
imply (\ref{sa}) \footnote{\label{foot1}To deduce
(\ref{sa}) from (\ref{san}, \ref{kia}, \ref{brox}) we
take in (\ref{san}) a convex function $f_m(y)={\rm
max}[y-\frac{\x_m(t)}{\p_m},0]$ and obtain (\ref{sa}) via 
$
{\sum}_{k=1}^m \p_k (\frac{\x_k(t+\epsilon)}{\p_k}-\frac{\x_m(t)}{\p_m}) 
\leq 
{\sum}_{k=1}^m \p_k f_m(\x_k(t+\epsilon)/\p_k) \leq{\sum}_{k=1}^n\p_k f_m(\x_k(t+\epsilon)/\p_k)
\leq {\sum}_{k=1}^n\p_k f_m(\x_k(t)/\p_k)
= {\sum}_{k=1}^m\p_k (\frac{\x_k(t)}{\p_k}-\frac{\x_m(t)}{\p_m}).
$}. Eq.~(\ref{san}) holds for all convex functions $f(y)$, if it holds for special classes of
convex functions; see Footnote \ref{foot1}. For an illustration of
(\ref{san}) take, e.g., $f(y)=|1-y|$. Then (\ref{san}) will demand
time-decay of ${\sum}_{k=1}^n|p_k-x_k(t)|$, which is an intuitive
measure of the perturbation magnitude. Many other such intuitive
measures are inlcluded in (\ref{san}) \cite{gorban}.

Eq.~(\ref{ra}) implies that a function $F[\bx(t);f]={\sum}_{k=1}^n p_k
f(x_k(t)/p_k)$ monotonously decays in time to its minimum $f(1)$, which
is reached for $\bx=\bp$. Convexity of $f$ implies that $f(1)$ is both local
and global minimum for $F[\bx(t);f]$. In contrast to asymptotic stability,
which always relates to some Lyapunov function \cite{chetaev}, the Le
Chatelier principle requires a specific {\it class} of Lyapunov
functions; see Footnote \ref{foot1} for one choice of such a class. 

\comment{
Among all those Lyapunov functions one often selects $f_{\rm
P}(x)=x\ln x$; see \cite{chakra} and references therein. The main reason
for this choice is that when the master equation (\ref{rota}) describes a
physical system coupled to a large quasi-equilibrium thermal bath, the
non-negative quantity $-\dot{F}[x(t);f_{\rm P}]={\sum}_{k=1}^n\dot{x}_k
\ln[p_k/x_k(t)]$ is linked to the {\it entropy production} in the
overall system [the system described by (\ref{rota}) plus the bath]
\cite{kampen}. For more general situations|when (\ref{rota}) is
applied to social or biological systems|insisting on the special role of
$f_{\rm P}(x)$ is not justified. Note that
$\dot{F}[x(t);f_{\rm P}]\leq 0$ does not guarantee the validity
of Le Chatelier principle (\ref{san}). Still from
$\dot{F}[x(t);f_{\rm P}]\leq 0$ one can deduce that the weak Le
Chatelier principle holds. 
}

Conditions (\ref{kia}, \ref{brox}, \ref{sa}) amount to the definition of
majorization order known in different areas of probability theory
\cite{ruch_mead,ruch_trio,olkin,ruch_lesche} and applied to master
equations in \cite{schlogl,alberti,michal,menon}: $\bx(t)$ majorizes 
\footnote{For a finite $\epsilon$, where (\ref{kia}) and (\ref{brox}) need not 
hold simultaneously, conditions (\ref{kia}, \ref{brox}, \ref{sa}) differ
from (\ref{san}). This is reflected by calling the former (latter) 
$p$-majorization ($d$-majorization) \cite{olkin}.
}
$\bx(t+\epsilon)$ relative to $\bp$ 
\BEA
\label{major}
\bx(t)\mgrt \bx(t+\epsilon), \qquad \epsilon\to 0+. 
\EEA
Note: {\it i)} $\bp\mgrt \bx$ leads to $\bx=\bp$, as follows from
(\ref{sa}) and the fact that (\ref{kia}) implies $1\leq
{\x_1(t)}/{\p_1}\geq {\x_n(t)}/{\p_n}\leq 1$. {\it ii)} $\bx(t)\mgrt \bp
$ for {\it any} $\bx(t)$. This follows from the convexity of $f(x)$ in
(\ref{san}) and means that being extended to long times, the Le
Chatelier principle becomes equivalent to asymptotic stability of $\bp$:
$\bx(t)\mgrt \bx(t+\infty)=\bp$. Hence one interpret (\ref{major}) in
terms of "information distance" \cite{ruch_lesche}: $\bx(t+\epsilon)$ is
closer to $\bp$ than $\bx(t)$. 

\subsection{Weak formulation of the Le Chatelier principle}
\label{weak}

When applying the Le Chatelier principle (\ref{sar}, \ref{kia},
\ref{brox}), it will prove convenient to separate the case where some
of conditions (\ref{sar}) [or (\ref{sa})] are violated, but not all of
them are violated simultaneously. This will be refered to as the weak
formulation of the Le Chatelier principle: although the perturbation is
not diminished, it is also not amplified. The weak formulation does not
imply asymptotic stability, but still it prevents strong unstabilities.
If {\it all} inequalities in (\ref{sar}) [or in (\ref{sa})] are reversed, we
get perturbation amplification that is in sharp contrast to the Le
Chatelier principle.  The weak formulation can be applied to stable [but
not asymptotically stable] rest-point. 

In the stability theory one typically studies consequences of a single
perturbation. It is recognized that the standard measures of stability
are insufficient for treating a sequence of multiple perturbations
\cite{justus}. The Le Chatelier principle treats this
situation, e.g. a violated weak principle would mean that a
sequence of weakening perturbations may drive the system towards the
boundaries of (\ref{ko}), where the extinction risks are sizable. 

Another way of weakening the formulation (\ref{sa}, \ref{sar}) of the Le
Chatelier principle is to require {\it only} decrease of the largest
component of the perturbation and increase of the smallest component:
$\dot{x}^\downarrow_1\leq 0$ and $\dot{x}^\downarrow_n\geq 0$. For
$n\geq 4$ this differs from (\ref{sar}). Likewise, a stronger [than
(\ref{sa},\ref{san})] formulation is also possible: one may require the
relation (\ref{sa}, \ref{kia}, \ref{brox}) to hold not only for
$\epsilon\to 0+$, but also for all $\epsilon>0$. Both are
worth exploring, but are not at focus here. 

\subsection{Linear master-equation}
\label{lin}

Eqs.~(\ref{san}, \ref{ra}) hold if in (\ref{rota}) $W_{ik}$ does not
depend on $\bx$ (linear master-equation)\cite{ruch_mead,ruch_trio,olkin,ruch_lesche,schlogl,alberti,michal,gorban}.
For a small $\epsilon$ and arbitrary $\bx(t)>0$ we write (\ref{rota}) as
$\bx(t+\epsilon)=\Pi \bx(t)$, where $\Pi$ is a stochastic matrix,
$\Pi_{ik}\equiv \delta_{ik}+\epsilon W_{ik}\geq 0$ and $\sum_{i=1}^n
\Pi_{ik}=1$, with $\Pi \bp=\bp$. Now (\ref{san}) follows after
representing its left-hand-side as
\BEA
\label{us}
{\sum}_{k=1}^np_k f\left(\frac{1}{p_k}{\sum}_{i=1}^n\Pi_{ki}p_i \,\,\frac{x_i(t)}{p_i}\right)
\EEA
and employing convexity of $f(x)$ via
$\frac{1}{p_k}{\sum}_{i=1}^n\Pi_{ki}p_i=1$. Eqs.~(\ref{san}, \ref{ra})
holds [via the same argument (\ref{us})] for the non-linear Boltzmann
master equation that describes equilibration of an closed macroscopic
system \cite{alberti}. Thus, for perturbations over thermodynamic equilibria we
confirmed the Le Chatelier principle {\it from within} non-equilibrium
statistical physics. 

Note however that the argument (\ref{us}) does not automatically apply
to all non-linear master-equations, since in general $W_{ik}[\bx]$ does
not leave invariant the rest-point $\sum_{k=1}^n W_{ik}[\bx]p_k\not =0$,
even if it is globally stable.

\comment{
\subsection{Structural perturbations}
\label{struc}

So far we considered a sudden perturbation of the state: $p\to x(t)$.
Certain perturbations do not act directly on the state $x$, but
influence the structure of the system making the transition probabilities
$W_{i\not=k}[x,\lambda_t]$ time-dependent via a 
parameter $\lambda_t$; see \cite{ives} for 
the relevance of such perturbations in ecology. Now the actual solution $x(t)$ of
(\ref{rota}) with $W_{i\not=k}[x,\lambda_t]$ is compared to the
case, where $\lambda_t$ is so slow that due to the assumed
asymptotic stability of the constant-$\lambda$ situation $x(t)$ remains
close to the {\it moving rest-point} $p(t)$ found from
\BEA
\label{kotosh}
{\sum}_{k=1}^n W_{ik}[p(t),\lambda_t]p_k(t)=0.
\EEA
Now the Le Chatelier principle is sought as $x(t+\epsilon)\prec_{p(t)} x(t)$
and means that the actual solution does not deviate too much from the moving 
rest-point. This holds for the linear master equation due to argument (\ref{us}).
The situation becomes less trivial if 
(\ref{kotosh}) does not admit proper solutions (or such solutions are not unique)
for some $\lambda_t$ along its trajectory.

}

\comment{
So far we considered a sudden perturbation of the state: $p\to x(t)$.
Certain perturbations do not act directly on the state $x$, but
influence the structure of the system: the interaction (pay-off)
parameters $a_{ik}$ are assumed to depend on time via a given
time-dependent parameter $\lambda_t$. This can model, e.g, environmental
influences on the replicator system \cite{armen_hu}. A theory for such
perturbations of the replicator equation was recently developed in
\cite{armen_hu}. 

Now the actual solution $x(t)$ of (\ref{rota}) with the time-dependent
$a_{ik}(\lambda_t)$ is compared to the situation, where $\lambda_t$ is
so slow that due to the assumed asymptotic stability of the
constant-$a_{ik}$ situation $x(t)$ remains close to the moving Nash
equilibrium $p(t)$ found from (\ref{laslo}) after $a_{ik}\to
a_{ik}(\lambda_t)$. The analogue of the representation (\ref{kara}) is
formulated as $a_{ik}(\lambda_t)=\widetilde{a}_{ik}(t)+c_{ik}(t)$, where
$\widetilde{a}_{ik}(t)$ supports $p(t)$ as the Nash equilibrium and
satisfies (\ref{merlin}), while $c_{ik}(t)$ supports two Nash equilibria
$p(t)$ and $x(t)$. If the so modified condition (\ref{kara}) holds, then
we still get the Le Chatelier condition (\ref{sa}, \ref{major}) with
$p\to p(t)$. Conditions (\ref{brat}) are generalized accordingly. 

Thus for structural perturbations the Le Chatelier principle is to be
formulated with respect to the moving Nash equilibrium and means that
the actual solution does not deviate too much from it. The situation
becomes less trivial and deserves a separate study whenever
$a_{ik}(\lambda_t)$ does not admit Nash equilibrium at some moments of
time or admits several Nash equilibria. }

\section{Replicator dynamics}
\label{repli}

\subsection{Equations of motion}

Evolutionary Game Theory (EGT) describes selection processes in a
population of interacting agents divided into groups [replicators] \cite{hofbauer}. The growth
of each group is governed by its fitness, which depends on inter-group interactions.
The frequency of the group $k$ is
$x_k={N_k}/{\sum_{k=1}^n N_k}$, where $N_k$ is the number of agents in
the group $k$. The simplest choice for the fitness $\phi_k$ of the group
$k$ 
\footnote{This does not imply that the approach necessarily refers 
to the so called group selection, although such a reference is not 
excluded; see \cite{mayr} for a recent careful discussion.}
is a linear function of frequencies \cite{hofbauer}:
\BEA
\phi_k[\bx]={\sum}_{l=1}^n a_{kl} x_l, \qquad k=1,\ldots, n, 
\EEA
where the {\it payoffs} $a_{kl}$ account for the interaction between
(the agents from) groups $k$ and $l$. The replicator equation 
\BEA 
\label{z2}
\dot{x}_i =  x_i (\phi_i[\bx]-{\sum}_{k=1}^n x_k\phi_k[\bx])
\EEA 
equates the growth per capita $\dot{x}_i/x_i$ to the fitness of group
$i$ minus the average fitness ${\sum}_{k=1}^n x_k\phi_k[x]$
\cite{hofbauer,sandholm}. Eq.~(\ref{z2}) is invariant with respect to
$a_{ik}\to a_{ik}+\widetilde{c}_k$, where $\widetilde{c}_k$ is any
constant. We set $\widetilde{c}_k$ such that $\{a_{kk}=0\}_{k=1}^n$. 

Solutions of (\ref{z2}) do not leave the domain (\ref{ko}); if
$x_k=0$ at an initial time it stays zero for all times.

Eq.~(\ref{z2}) can be interpreted via a {\it symmetric} game played by
two participants {\it I} an {\it II} \cite{hofbauer,sandholm}. Here the
groups correspond to strategies of the play, while $a_{ik}$ ($a_{ki}$)
is the pay-off received by {\it I} ({\it II}) when applying strategy $i$
against $k$ ($k$ against $i$). $x_i$ is the probability of applying the
strategy $i$; $\{x_k\}_{k=1}^n$ do not depend on the player, since the
game is symmetric. Hence $\sum_{k=1}^n a_{ik}x_k$ is the average pay-off
received by {\it I} in response to applying the strategy $i$
\cite{hofbauer}. Now (\ref{z2}) refers to the feedback between applying
a strategy and the pay-off received for it \cite{hofbauer,sandholm}. 

Eq.~(\ref{z2}) can be derived from the non-linear master
equation (\ref{rota}) under suitable choice of $W_{i\not =k}[\bx]$
\cite{hofbauer,sandholm}:
\BEA
\label{grimm}
W_{i\not =k}[\bx]=x_i({\cal K}+\phi_i[\bx]-\phi_k[\bx])/2,
\EEA
where ${\cal K}$ is a constant ensuring $W_{i\not =k}[\bx]\geq 0$
(transition probability is non-negative). $W_{kk}[\bx]$ is found
from normalization ${\sum}_{i=1}^n W_{ik}[\bx] =0$. Hence
the transition probability $W_{i\not =k}[\bx]$ of an agent from
group $k$ to $i$ is facilitated by the probability $x_i$ of that
group and the fitness difference. The choice (\ref{grimm}) is not
unique. Other options, e.g., $W_{i\not =k}[\bx]=x_i({\cal
K}_1+\phi_i[\bx])$, have a different meaning, but lead to the same equation.

The meaning of $a_{k\not =l}$ is best illustrated by
rewriting (\ref{z2}) via $\{N_k\}_{k=1}^n$:
\BEA
\label{polo}
\dot{N}_i=N_i{\sum}_{k=1}^na_{ik}x_k, \quad x_k=\frac{N_k}{N}, \quad N={\sum}_{l=1}^n N_l. ~~
\EEA
Now $a_{k\not =l}=\frac{\delta [\dot{N}_k/x_k]}{\delta N_l}$ describes the change of the relative
growth of the group $k$ due to a small change $\delta N_l$ in the group
$l$.  Recalling that $a_{ll}=0$, we have four possibilities for the
interaction between $k$ and $l$ \cite{svi}:
\BEA
\label{tutu1}
&&a_{k\not =l} >0, ~~ a_{l\not =k} >0: ~~ {\it mutualism} \\
\label{tutu2}
&&a_{k\not =l} <0, ~~ a_{l\not =k} <0: ~~ {\it competition} \,~~~~\\
\label{tutu3}
&&a_{k\not =l} >0, ~~ a_{l\not =k} <0: ~~ k\,\,{\it predates } \,\,l. 
\EEA
Fourth possibility is that $l$ predates $k$. 

For the game-theoretical
meaning of (\ref{tutu1}, \ref{tutu2}) recall that we consider a
symmetric game, where $\{a_{lk}\}$ and $\{a_{kl}\}$ are the pay-off
matrices for players {\it I} and {\it II}, respectively. Hence 
$a_{ll}>a_{k\not =l}$ and $a_{ll}>a_{l\not =k}$ (competition) means that 
{\it I} and {\it II} will tend to coordinate their actions. Likewise,
mutualism relates with anticoordination. 

Applications of replicator dynamics include: {\it i)} Sociobiology,
where the replicators correspond to the strategies of agent's behavior,
while $x_k$ is the probability by which an agent applies the strategy
$k$ \cite{hofbauer}; $x_k$ can change due to inheritance, learning,
imitation, infection, {\it etc}. {\it ii)} Genetic selection, where
$x_k$ is the frequency of one-locus allele $k$ in panmictic, diploid
population, and where $a_{kl}=a'_{kl}+a''_k$ combines the selective
value $a'_{kl}=a'_{lk}$ of the phenotype driven by the zygote $(kl)$ and
the selective value $a''_{k}$ of the gamete $(k)$ \footnote{Note basic
limitations of this selection model \cite{passekov,poluektov,hofbauer}:
the recombination is ignored; differences between the sexes are ignored,
e.g. one ignores the fact that the female and male gametes can have
different selective values; it is assumed that the one-locus gene does
not determine the sex.} \cite{passekov,poluektov,hofbauer}.
Eq.~(\ref{z2}) with such a pay-off matrix $\{a_{kl}\}$ was proposed in
30's within population genetics, where it describes selection dynamics
\cite{passekov,poluektov,hofbauer}. {\it iii)} Lotka-Volterra equations
of ecological dynamics \cite{hofbauer}. {\it iv)} Quantum feedback
control, where the replicator equation (\ref{z2}) comes out from the
Schroedinger equation and the pay-offs are anti-symmetric (zero-sum
game): $a_{kl}=-a_{lk}$ \cite{armen_mahler}. {\it v)} Solution of hard
combinatorial optimization problems and genetic algorithms \cite{menon}. 

\subsection{Nash equilibrium}
\label{dbala}

If (\ref{z2}) admits a rest-point $\bp^{\rm T}=\{p_k>0\}_{k=1}^n$, then 
\BEA
\label{laslo}
{\sum}_{k=1}^n a_{ik}p_k=\phi,\quad p_i=\frac{\sum_{k=1}^n(a^{-1})_{ik}}{\sum_{l,m=1}^n(a^{-1})_{lm}},
\EEA
where the fitness $\phi$ does not depend on $i$: coexisting
groups are equally fit. If the rest point $\bp>0$ exists, it is unique among all
probability vectors with strictly positive components, since generically (modulo
small changes in $a_{ik}$) the matrix $A=\{a_{ik}\}_{i,k=1}^n$ is invertible \cite{zeeman}. Due to
(\ref{laslo}) the Nash equilibrium condition holds for $\bp$ [$\bq$ is any
probability vector]
\BEA
{\sum}_{i,k=1}^n q_ia_{ik}p_k\equiv \bq^{\rm T}A\bp\leq \bp^{\rm T}A\bp
\label{nash}
\EEA
with the equality sign \cite{hofbauer}. Eq.~(\ref{nash}) means that a
player applying the strategies with probability vector $\bp$ does not get
incentives for a unilaterial change $\bp\to \bq$. A Nash equilibrium need
not be a stable rest point of the replicator dynamics \cite{zeeman,hofbauer,sandholm}. 

The fact that at a Nash equilibrium coexisting groups are equally fit can be
related to the detailed balance, a basic notion of the
thermodynamic equilibrium \cite{kampen}. A stationary state of
(\ref{rota}) means: $\dot{p}_k=0$ for $k=1,\ldots,n$.  Since
(\ref{rota}) can be written as $\dot{p}_i = {\sum}_{k\not =i}^n
(W_{ik}[\bp]p_k-W_{ki}[\bp]p_i)$, an additional feature of such a stationary
state is that
\BEA
\label{db}
W_{ik}[\bp]p_k=W_{ki}[\bp]p_i~~~ {\rm for~ all}~~ i\not=k, 
\EEA
i.e. transitions from one group to another are in the detailed balance
\cite{kampen}. This condition holds for the thermodynamic
equilibrium, where it relates to the time-inversion invariance of
microscopic (Hamiltonian) equations of motion \cite{kampen}. Now
(\ref{grimm}) and (\ref{laslo}) imply that a Nash equilibrium also
satisfies the detailed balance condition, one reason to regard it
as an extension of the thermodynamic equilibrium. 

The satisfaction of the detailed balance was noted for a class of other
game-theoretical models \cite{bertin}. The authors of \cite{bertin}
also relate the game-theoretical fitness to the chemical potential.
There is a clear analogy here: surviving species have equal fitness at a
Nash equilibrium, while chemical potentials of interacting systems are equal 
in the thermodynamic equilibrium. 

\subsection{Evolutionary stability}
\label{ESS}

\subsubsection{Standard viewpoint}

We recall the concept of evolutionary stability, which (for symmetric
games) refines the Nash equilibrium and is widely involved in
applications \cite{zeeman,hofbauer,sandholm}. Below we see how the Le
Chatelier principle generalizes this notion. 

A Nash equilibrium (\ref{laslo}, \ref{nash}) is evolutionary stable if for any
probability vector $\bx>0$ \cite{hofbauer}
\BEA
\bx^{\rm T}A\bx-\bp^{\rm T}A\bx =(\bx^{\rm T}-\bp^{\rm T})A(\bx-\bp)<0,
\label{ess}
\EEA
where in the last expression we used $(\bx^{\rm T}-\bp^{\rm T})A\bp=0$. 
It follows from the fact that $\sum_{k=1}^n a_{ik}p_k$ does not depend
on the index $i$ [see (\ref{laslo})] and that $\sum_{k=1}^n (x_k-p_k)=0$.

The game-theoretical meaning of (\ref{ess}) is that albeit the Nash
condition (\ref{nash}) holds with equality for any $\bx\not =\bp$, such $\bx$
cannot become a Nash equilibrium, since $\bp$ fairs better against any
$\bx$. Eqs.~(\ref{ess}, \ref{z2}) imply the Lyapunov feature for $g(\bx,\bp)={\sum}_{k=1}^np_k
\ln\frac{p_k}{x_k(t)}$:
\BEA
\frac{\d g(\bx,\bp)}{\d t}= \bx^{\rm T}A\bx-\bp^{\rm T}A\bx<0,~
\label{naga}
\EEA
and $g(\bx,\bp)\geq 0$ with $g(\bx,\bp)=0$ if and only if $\bx=\bp>0$. Evolutionary
stability implies global stability, because due to (\ref{nash}, \ref{ess}) there are
no other internal Nash equilibria. In this sense 
evolutionary stability refines the notion of the Nash equilibrium, which
need to be neither unique nor stable. But the 
evolutionary stability does not extend to asymmetric games, since they do not have
asymptotically stable interior Nash equilibria \cite{hofbauer}.

Since $g(\bx,\bp)$ is a convex function of $\frac{x_k(t)}{p_k}$, one cannot
get all the inequalities in (\ref{sa}) [or (\ref{san})] reversed. Hence
evolutionary stability implies the weak Le Chatelier principle. The
converse is not true: there are situations, where the asymptotically
stable Nash equilibrium is not evolutionary stable, but the weak Le
Chatelier principle holds [see below].  The necessary and sufficient
condition for an interior Nash equilibrium $p$ to be evolutionary stable is deduced from (\ref{ess})
and $\sum_{i=1}^n(x_i-p_i)=0$: the matrix 
\BEA
\label{symo}
\{\, a^{[s]}_{in}+a^{[s]}_{kn}-a^{[s]}_{ik}\,\}_{i,k=1}^{n-1}
\EEA
is positive definite. Here $a^{[s]}_{ik}\equiv\frac{1}{2}[a_{ik}+a_{ki}]$. 

\comment{
There are important cases, where the Nash equilibrium is dynamically
relevant, but it is not asymptotically stable thus trivially excluding
the evolutionary stability: (1) for asymmetric (two-population) games
internal Nash equilibria cannot be asymptotically stable, they are
stable at best \cite{hofbauer}. (2) replicators are {\it permanent}, if
no internal trajectory converges to the simplex boundary
\cite{hofbauer}. The permanence generalizes the concept of stability,
since all initially present groups survive for all times.  Then there is
a {\it unique} internal Nash equilibrium $p$, which is generally {\it
not} stable, but which nevertheless governs the long-time dynamics via
$\int_0^T\frac{\d t}{T}\,x_i(t)\to p_i$ for $T\to\infty$
\cite{hofbauer}. As seen below, for such cases one can still apply the
Le Chatelier principle and its weak version. }

\subsubsection{Mutualism in the whole and evolutionary stability}

The evolutionary stability can be given a more general 
meaning related to the population being {\it mutualistic in the whole}
\cite{poluektov}. Eqs.~(\ref{z2}, \ref{polo}) show that the overall
number of agents $N={\sum}_{k=1}^n N_k$ in the population evolves as
$\dot{N}=N{\sum}_{k=1}^na_{ik}x_kx_i\equiv N\phi[\bx]$, where $\phi[\bx]$ is
the average fitness. Note that $N$ and $\bx$ are independent variables. 

Let there be another population with the same pay-offs $\{a_{kl}\}$, but
with different parameters $\bx'$ and $N'$ (frequencies and the overall
number of agents). If these two polulations are joined together, the
number of agents in each group $k$ becomes $N_k+N_k'$. The speed of the
overall number of agents in the joint population is $(N+N')\phi[\lambda
x+(1-\lambda)x']$, where $\lambda=\frac{N}{N+N'}$, $x_k=\frac{N_k}{N}$
and $x'_k=\frac{N'_k}{N'}$. The mutualism in the whole demands that the
speed of the overall number of agents in the joint population is larger
than the sum of separate speeds \cite{poluektov}:
\BEA
\label{golem}
\lambda \phi[\bx]+(1-\lambda)\phi[\bx']\leq \phi [\lambda \bx+(1-\lambda)\bx'].
\EEA
Since $0<\lambda<1$, (\ref{golem}) amounts to the average fitness
$\phi[\bx]$ being a concave function of $\bx$ on the simplex ${\bf S}_n$
\cite{poluektov}. Let now (\ref{golem}) holds for any two probability
vectors $\bx$ and $\bx'$.  Assuming $\bx\approx \bx'$ in (\ref{golem}) and
expanding it over a small $\bx-\bx'$, we get $\bq^{\rm T}A\bq\leq 0$ for any
vector $\bq$ with ${\sum}_{k=1}^n \bq_k=0$. This is condition
(\ref{ess}) for the evolutionary stability. Note that the mutualism in
the whole does not mean that separate inter-group interaction are also
mutualistic in the sense of (\ref{tutu1}). {\it Vice versa}, the mutualistic 
interactions do not imply the mutualism in the whole; see below for a 
concrete example. The Le Chatelier principle
implies the mutualism in the whole due to (\ref{naga}). 

\section{Criteria of the Le Chatelier principle for replicator dynamics}
\label{crit}

\subsection{Criteria for a given perturbation}

Below we determine conditions under which the replicator equation
(\ref{z2}) leads to the Le Chatelier principle (\ref{sa}, \ref{sar})
with respect to the rest point $\bp>0$. We assume that perturbations
satisfy $\bx(t)>0$. Hence we exclude invasive perturbations,
where some of $p_k$'s nullify, and extinctions, where some of
$x_k(t)$'s nullify. 
Write (\ref{z2}) as 
\BEA 
\label{rota_r}
\dot{x}_i  = {\sum}_{k=1}^n V_{ik}[\bx]x_k,~ V_{ik}[\bx] \equiv x_i(a_{ik}-{\sum}_{l=1}^nx_la_{lk}).\,\,
\EEA 
Generally, $V_{i\not =k}[\bx]$ is not a transition probability, but it leaves intact
an internal Nash equilibrium $\bp>0$:
\BEA
{\sum}_{k=1}^n V_{ik}[\bx]p_k=0,
\label{pin}
\EEA
because $\phi$ in (\ref{laslo}) does not depend on the index $i$.

Let for certain pay-offs $\widetilde{a}_{ik}$ 
the corresponding $\widetilde{V}_{i\not =k}$ be positive for all pairs $i\not =k$
\BEA
\label{merlin}
\widetilde{V}_{i\not = k}[\bx]\geq 0 ~~~~{\rm or}~~~~ \widetilde{a}_{i\not =k}\geq {\sum}_{l=1}^n x_l \widetilde{a}_{lk}.
\EEA
Then $\Pi_{ik}[\bx]=\delta_{ik}+\epsilon \widetilde{V}_{ik}[\bx]$ is a
stochastic matrix with $\Pi\bp=\bp$ due to (\ref{pin}). Hence (\ref{us})
applies and (\ref{merlin}) suffices for the validity of the Le Chatelier
principle (\ref{sa}, \ref{sar}). 

If now (\ref{merlin}) holds for at least one $\bx>0$ (with $\sum_{k=1}^nx_k=1$),
it leads to ${\sum}_{k=1}^n x_l \widetilde{a}_{lk}\geq 0$ (recall that $\widetilde{a}_{kk}=0$),
which combined with (\ref{merlin}) produces
\BEA
\label{gla}
\widetilde{a}_{i\not =k} \geq 0 ~~{\rm ~for~ all~ pairs}~~~ i\not =k,
\EEA
i.e., mutualism or anticoordination; see (\ref{tutu1}, \ref{tutu2}).

Let us now assume that the principle holds for a perturbation $\bx(t)$: $\bx(t)\mgrt
\bx(t+\epsilon)$. Then it is shown in Ref.~\cite{ruch_trio} that
there exists a stochastic matrix|which for our
purposes can be taken as $\Pi_{ik}[\bx]=\delta_{ik} +\epsilon
\widetilde{V}_{ik}[\bx]$ with $\widetilde{V}_{i\not =k}[\bx]\geq 0$|such that
$\Pi[\bx] \bp=\bp$ and $\bx(t+\epsilon)=\Pi[\bx]\bx(t)$ or 
$\dot{x}_i(t)=\sum_{k=1}^n \widetilde{V}_{ik}[\bx] x_k(t)$. 

We define from (\ref{rota_r}) the corresponding
pay-off matrix: $\widetilde{a}_{ik}[\bx]=\frac{\widetilde{V}_{ik}[\bx]}{x_i}-\frac{\widetilde{V}_{kk}[\bx]}{x_k}$,
where $\widetilde{a}_{kk}[\bx]=0$ due to $\sum_{i=1}^n
\widetilde{V}_{ik}[\bx]=0$. Now $\widetilde{a}_{ik}[\bx]$ admits the Nash
equilibrium $\bp$ and satisfies (\ref{merlin}). Since $a_{ik}$
and $\widetilde{a}_{ik}[\bx]$ generate the same local flow, 
the pay-off matrix $c_{ik}[\bx]\equiv a_{ik}-\widetilde{a}_{ik}[\bx]$
admits another Nash
equilibrium $\bx(t)$ besides $\bp$; see (\ref{nash}). 
Hence, ${\rm det}\,[c_{ik}]=0$; see (\ref{laslo}). 

Thus, the necessary and sufficient condition for the Le Chatelier
principle (\ref{sa}, \ref{san}) to hold with respect to a perturbation
$\bx(t)>0$ over the internal Nash equilibrium $\bp>0$ is that the pay-off matrix
$a_{ik}$ is represented as 
\BEA
a_{ik}=\widetilde{a}_{ik}[\bx]+c_{ik}[\bx],
\label{kara}
\EEA
where $\widetilde{a}_{ik}[\bx]$ satisfies (\ref{merlin}) and has the Nash
equilibrium $\bp$, while the matrix $c_{ik}[\bx]$ has two Nash equilibria,
$\bx(t)$ and $\bp$.  Using (\ref{rota_r}) we get the analogue of
(\ref{kara}) for generators:
\BEA
\label{mara}
V[\bx] =\widetilde{V}[\bx]+\Gamma[\bx],~~~
\widetilde{V}[\bx]\bp=\Gamma[\bx] \bp=\Gamma[\bx] \bx(t)=0,~
\label{marat}
\EEA
where $\Gamma_{ik}[\bx]=x_i(c_{ik}[\bx]-{\sum}_{l=1}^nx_lc_{lk}[\bx])$, and
where $\widetilde{V}[\bx]$ satisfies (\ref{merlin}).
Eq.~(\ref{mara}) refers to hidden (by $\Gamma[\bx]$) {\it mutualism} (or anti-coordination).
\comment{Eqs.~(\ref{kara}, \ref{marat}) are useful for generating examples, but we lack general conditions of
whether a matrix $a_{ik}$ can be represented as in (\ref{kara}). }

The weak Le Chatelier principle holds for a perturbation $\bx(t)$, if $\bx(t)\not\prec_p
\bx(t+\epsilon)$ [$\epsilon\to +0$]. Now $\bx(t+\epsilon)=(1+\epsilon
V[\bx])\bx(t)$ leads to $\bx(t)=(1-\epsilon V[\bx])\bx(t+\epsilon)$, so that the criteria for
the weak principle is read-off from (\ref{marat}): 
\BEA
\label{karpaty}
V[\bx]\neq-\widetilde{V}[\bx]+\Gamma[\bx], \qquad \widetilde{V}_{i\not=k}[\bx]\geq 0,
\EEA
where $\Gamma[\bx]$ satisfies the same conditions as in (\ref{mara}).
Eqs.~(\ref{karpaty}, \ref{tutu1}, \ref{tutu2}) show that violations of
the weak principle relate to hidden (by $\Gamma[\bx]$) {\it competition}
(coordination).

\subsection{Local validity of the principle}

Local means that we consider small perturbations:
$x_i(t)-p_i\ll p_i$. Studying local perturbations is important, since
most of perturbations arising due to internal noise will be small.
Another reason is that locally (around an asymptotically stable interior
rest-point) many different dynamic evolutionary approaches are
equivalent \cite{hofbauer}.  We linearize (\ref{z2}, \ref{rota_r})
around $\bp$:
\BEA
\label{orsk}
\dot{\xi}_i = {\sum}_{k=1}^n V_{ik}[\bp]\xi_k, \quad \xi_i\equiv x_i-p_i,
\EEA
where we used (\ref{laslo}). The local stability is governed by the
matrix $V_{ik}[\bp]$. The determinant of $V_{ik}[\bp]$ (and thus at least
one of its eigenvalues) is zero, because $\sum_{i=1}^n V_{ik}[\bp]=0$. 
All other eigenvalues of $V_{ik}[\bp]$ have negative real part, because $\bp$ is assumed to be
asymptotically stable. 

The criteria for the local validity of the Le Chatelier principle amount
to changing $\bx\to \bp$ in (\ref{mara}). Alternatively, we can proceed
directly from (\ref{orsk}) and demand validity of (\ref{major}) for
$\xi(j)=\varepsilon[\bx(j)-\bp]$, where $\varepsilon \ll 1$ is a small
parameter, and $\bx(j)^{\rm T}=\{\delta_{jk}\}_{k=1}^n$. We then conclude that the
principle holds for all local perturbations if and only if
$V_{i\not=k}[\bp]\geq 0$, or
\BEA
\label{baran}
a_{i\not =k}\geq {\sum}_{l=1}^n p_l a_{lk} ~~{\rm ~for~ all~ pairs}~~~ i\not =k.
\EEA
Note that mutualism ($a_{i\not =k}\geq 0$) follows from (\ref{baran}),
but does not suffice for the validity of the Le Chatelier principle for
all local perturbations. The game-theoretic meaning of (\ref{baran}) is
that when responding to a pure strategy $k$, it is better to use 
{\it any other pure} strategy $i$ than to respond via the Nash equilibrium mixed
strategy $\bp$. 

Eq.~(\ref{ra}) implies that if the principle holds for all local perturbations, and
if $f(x)$ is smooth [recall $f''(1)>0$],
\BEA
\label{leo}
\frac{\d}{\d t}{\sum}_{k=1}^n\frac{\xi_k^2}{p_k}\leq 0.
\EEA
The converse is not true: though for local perturbations (\ref{leo})
does not depend on the form of $f$ for a class of functions $f(x)$, it
does not suffice for the local validity of the Le Chatelier principle
[recall that the class of functions in Footnote \ref{foot1} is not
smooth].  Put differently, the local Le Chatelier principle {\it is not}
a consequence of the evolutionary stability that also produces
(\ref{leo}) locally; cf. (\ref{leo}, \ref{naga}) with (\ref{baran}) and
see section \ref{mutu} for an explicit example. 

By analogy to phenomenological non-equilibrium thermodynamics,
(\ref{leo}) is sometimes presented as the statement of the Le Chatelier
principle for population dynamics; see \cite{chakra} and references
therein. We stress that (\ref{leo}) is only a necessary condition for
the local principle. 

We have the following interplay of the mutualism concepts and their
relation to the Le Chatelier principle: the mutualism in the whole
(evolutionary stability) does not imply (\ref{tutu1}), but prevents weak
violations of the principle. The validity of the local principle
requires conditions (\ref{baran}) which are stronger than the mutualism
in the whole.

\comment{
Le Chatelier principle can now be adapted to this local situation. 
Given a vector $\xi=(\xi_1,\ldots,\xi_n)$ with 
\BEA
\label{ura}
\xi_1/p_1> \xi_2/p_2>\ldots \xi_n/p_n~~{\rm and~with}~~{\sum}_{i=1}^n \xi_i=0, 
\EEA
the local validity of the Le Chatelier principle means
\BEA
V[p]\,\xi^{\rm T}\mless (0,0,\ldots,0)^{\rm T}\equiv 0.
\label{gora}
\EEA
For (\ref{gora}) to hold for all $\xi$'s from (\ref{ura}) [i.e., for all
local perturbations] it is necessary and sufficient that conditions
(\ref{merlin}) hold with $x=p$. This can be established in two ways.  We
can note from the discussion before (\ref{kara}) that $c_{ik}=0$ is the
only pay-off matrix having every points in a vicinity of $p$ as its Nash
equilibrium.  Alternatively, we can employ (\ref{gora}) for all those
$\xi=x-p$, where $x$ has one of its components equal to $1$, while all
other components are zero. Likewise, the Le Chatelier principle is
violated if for at least one $\xi=\eta\not=0$ from (\ref{ura}) the
majorization relation (\ref{gora}) is inverted. Then the matrix $V[p]$
can be represented as
\BEA
\label{brat}
V[p] =\widetilde{V}+\Gamma, ~~{\rm where}~~\widetilde{V}p^{\rm T}=\Gamma p^{\rm T}=\Gamma \eta^{\rm T}=0,
\EEA
and where $\widetilde{V}$ satisfies conditions (\ref{merlin}) with $x=p$.
}

\section{The Le Chatelier principle for $n=3$}
\label{n3}

\subsection{Global validity of the principle}

Note that $n=3$ (three groups) is the simplest non-trivial situation,
because for $n=2$ the Le Chatelier principle coincides with the notion
of asymptotic stability: an internal asymptotically stable state exists
only for mutualistic interactions, where (\ref{baran}) holds trivially. 

\begin{figure}
\vspace{0.2cm}
\includegraphics[width=8.7cm]{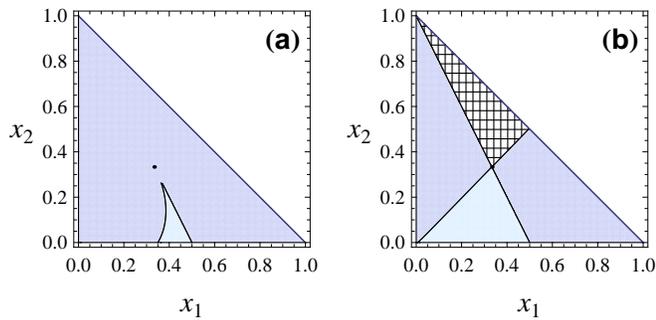}
\vspace{0.3cm}
\caption{(Color online.)
The structure of (\ref{z2}) on the simplex (\ref{ko})
[$0<x_1$, $0<x_2$, $0<x_1+x_2<1$] containing globally stable Nash
equilibrium (\ref{bask}) (bold dot) for $n=3$ and $\vartheta=1$; see
(\ref{nola}). Fig.~1(a): $b_1=b_2=b_3=0.3$. Fig.~1(b):
$b_1=b_2=b_3=0.99$. The Le Chatelier principle supporting domain is shaded normally (blue).
Light (blue) shaded domains support only the weak principle, where both strongest and weakest components of a perturbation grow.
Meshed domains also support only the weak principle, but now both strongest and weakest components of a perturbation decrease.
Such domains are absent for the situation presented on Fig.~1(a).
The Nash equilibrium on Fig.~1(b) is not evolutionary stable, but still globally stable. 
} 
\label{f2}
\end{figure}

Eq.~(\ref{merlin}) for $n=3$ reduces to 
\footnote{
Resolving these inequalities is a straightforward task in linear
programming; here is one example of solutions of
(\ref{kuk}): $\wa_{12}\leq \wa_{32}$, $\wa_{21}\leq
\wa_{31}$, $\wa_{13}\leq \wa_{23}$, $\frac{x_1}{1-x_2}\geq
1-\frac{\wa_{21}}{\wa_{31}}$, $\frac{\wa_{13}}{\wa_{23}}\geq
\frac{x_2}{1-x_1}\geq1-\frac{\wa_{21}}{\wa_{31}}$.  This is a triangle
embedded into the simplex (\ref{ko}). }. 
\BEA
\label{kuk}
{\rm min}[\wa_{k\not=i},\wa_{j\not=i}]\geq x_k\wa_{k\not=i}+x_j\wa_{j\not=i},~~ i,j,k=1,2,3.~
\EEA
\comment{\BEA
\label{kuk1}
{\rm min}[\wa_{12},\wa_{32}]\geq x_1\wa_{12}+x_3\wa_{32},\\
\label{kuk2}
{\rm min}[\wa_{21},\wa_{31}]\geq x_2\wa_{21}+x_3\wa_{31},\\
{\rm min}[\wa_{13},\wa_{23}]\geq x_1\wa_{13}+x_2\wa_{23}.
\label{kuk3}
\EEA}
For the Le Chatelier principle (\ref{sar}) to hold
for the whole simplex (\ref{ko}), we need $c_{ik}=0$ in (\ref{kara}), and
(\ref{kuk}) produces:
\BEA
a_{12}= a_{32}\geq 0,  ~~~ a_{21}=a_{31}\geq 0, ~~~ a_{13}= a_{23}\geq 0.
\label{hmhm}
\EEA
Hence the global validity of the Le Chatelier principle relates to
symmetrically mutualistic pay-offs. Under less restrictive
assumptions one can still get the principle valid for the most of
perturbations; see Fig.~\ref{f2} and discussion below. Moreover, for
$n=3$ we observed that an interior, stable Nash equilibrium always has
perturbation domains, where the Le Chatelier principle holds, and that
these domains include the Nash equilibrium, or the latter is contained
at their boundary; see Figs.~\ref{f2}--\ref{f3}. Recall that the
converse is guaranteed: the Le Chatelier principle implies asymptotic
stability, since it is related to the existence of a family of Lyapunov
functions (\ref{ra}). 

\subsection{Barocentric parametrization }

Below we study examples displaying concrete scenarios for (in)validity
of the principle.  For the sake of illustration we make another
simplification assuming that the asymptotically stable Nash equilibrium
is barocentric
\footnote{\label{barobaro}Assuming that there is a Nash equilibrium $p>0$, one can make the barocentric transformation:
$x_i'=\frac{x_i}{p_i}/[\sum_{k=1}^3\frac{x_k}{p_k}]$, which
maps the original replicator equation to the one with variables
$x_i'$ and payoff coefficients $a_{ik}'=a_{ik}p_k$
\cite{zeeman}. The Nash equilibrium in the new representation coincides with (\ref{bask}).
The barocentric
transformation respects stability features of a rest-point \cite{zeeman}, 
though the conditions of the Le Chatelier principle [as well as the conditions for evolutionary stability] are not respected.
However, due to $\frac{\dot{x}_i'}{x_i'}=\frac{\dot{x}_i}{x_i}-\frac{\d}{\d t}\ln [\sum_k \frac{x_k}{p_k}]$, 
if the principle holds in coordinates $x$, it at least weakly holds in coordinates $x'$.}:
\BEA
\label{bask}
p_1=p_2=p_3={1}/{3}.
\EEA
Then the pay-off matrix can be parametrized as
\BEA
\left(\begin{array}{rrr} 
0~~ & ~a_{12} & ~a_{13} \\ 
a_{21} & ~0~~ & ~a_{23}  \\
a_{31} & ~a_{32} &~0~~ 
\end{array}\right)=
\left(\begin{array}{rrr} 
0~~~ & \vartheta+b_1 &\vartheta-b_1 \\ 
\vartheta+b_2 & 0~~~ & \vartheta-b_2  \\
\vartheta+b_3&\vartheta-b_3 &0~~~ 
\end{array}\right), 
\label{nola}
\EEA
where $\vartheta$ and $b_i$ are real parameters.
Putting (\ref{nola}) into (\ref{orsk}) we get for two non-zero eigenvalues of $V_{ik}[\bp]$:
\BEA
[\,-\vartheta\pm\sqrt{b_1b_2-b_1b_3+b_2b_3}\,]/3. 
\label{ghost}
\EEA
Third eigenvalue of $V_{ik}[\bp]$ is zero. For asymptotic stability the real parts of
(\ref{ghost}) have to be negative; hence $\vartheta>0$. We can take $\vartheta=1$, since the
magnitude of $\vartheta$ can be rescaled with the characteristic times.

The barocentric condition (\ref{bask}) means that the Nash equilibrium
is maximally mixed and that the (minus) entropy $\sum_{k=1}^n x_{k}(t)\ln
x_k(t)$ is included in the class (\ref{san}) of Lyapunov functions.

\comment{
For (\ref{bask}) the matrix $\G$ from (\ref{mara}, \ref{marat})
has two linearly-independent right eigenvectors refering to the
eigenvalue zero: $(1,1,1)^{\rm T}$ (which is both left and right
eigenvector of $\G$) and $x(t)^{\rm T}$. Hence
$\G$ can be written as $\G=\xi^{\rm T}\otimes \eta$, 
\begin{gather}
\label{romo}
\G =
\left(\begin{array}{rrr} 
\xi_1\eta_1 & \xi_1\eta_2 & \xi_1\eta_3 \\ 
\xi_2\eta_1 & \xi_2\eta_2 & \xi_2\eta_3  \\
\xi_3\eta_1 & \xi_3\eta_2 & \xi_3\eta_3 
\end{array}\right),~
\sum_{k=1}^3\xi_k=\sum_{k=1}^3\eta_k=0,
\end{gather}
where $\xi^{\rm T}$ and $\eta$ are, respectively, left and right
eigenvectors of $\G$ referring to its non-zero eigenvalue.  Conversely,
any two vectors $\xi$ and $\eta$ satisfying (\ref{romo}) imply $\Gamma
x(t)^{\rm T}=0$ for some probability vector $x(t)$ linearly independent
from (\ref{bask}). Eq.~(\ref{romo}) can be used for generating examples
supporting or violating the principle. }

\subsection{Mutualistic replicators}
\label{mutu}

They are defined by condition $a_{ik}\geq 0=a_{kk}$. An internal Nash
equilibrium need not exist for mutualistic replicators, but if exists,
then (at least for $n=3$) it is globally stable; cf. (\ref{ghost}) under
$|b_i|\geq 1$ and Footnote \ref{barobaro}. 

Let us now show that for mutualistic replicators the weak principle
holds, at least for $n=3$. Recall (\ref{karpaty}) and the discussion
after it. If the weak principle were violated there would exist a
$3\times 3$ matrix $a_{ik}+\widetilde{a}_{ik}$ whose non-diagonal
(diagonal) elements are positive (zero), and whose determinant is zero
(since it supports two internal [linearly independent] Nash equilibria).
It is clear that such a matrix does not exist \footnote{We believe this
argument holds more generally, but we were not able to formalize it for $n\geq 4$.}.

For the case (\ref{nola}) the mutualism (\ref{tutu1}) amounts to $|b_i|\leq \vartheta=1$.
For $n=3$ the criteria of evolutionary stability reduce from
(\ref{symo}) to 
\BEA
\label{orme}
a_{13}^{[s]}\geq 0, \qquad 4a_{13}^{[s]}a_{23}^{[s]}\geq (a_{13}^{[s]}+a_{23}^{[s]}-a_{12}^{[s]})^2.
\EEA
To illustrate various stability notions, let us assume that
$\vartheta=1>b_i=b>0$.  Now (\ref{ghost}) implies that the Nash
rest-point (\ref{bask}) is asymptotically stable. The evolutionary
stability (\ref{orme}) is more restrictive, since it demands
$b<\sqrt{3}/2$. The validity of the Le Chatelier principle for all local
perturbations is even more restrictive, since it demands $b<1/3$; see
(\ref{baran}) and Fig.~\ref{f2}. Recall that the
weak principle always holds for mutualistic replicators. Hence the weak
principle is more general than the evolutionary stability, and the
latter does not imply the local validity of the principle. 

Note that the weak Le Chatelier principle has the following two
scenarios of validity: {\it i)} perturbations leading to growth, because
both the weakest and the strongest component of the perturbation
increase; {\it ii)} perturbations leading to decay, since now both the
weakest and the strongest component decrease. Other types of
perturbations for $n=3$ will either violate the weak principle or satisfy
the proper principle. It should be clear that during the second scenario
the system in a sense destablizes, because smaller sizes of groups means
larger extinction risks. It is seen from Figs.~\ref{f2} and \ref{f3}
that generally both scenarios for the weak Le Chatelier principle are
present. However, we observed that whenever for mutualistic replicators
the principle holds for all local perturbations, the domains that
support only the weak principle are exclusively of the first type: they
suppport the growth of population and not its decay; see
Fig.~\ref{f2}(a). This observation indicates that the local condition (\ref{baran}) 
does a play a certain global role as well.

\comment{
Returning to the representation (\ref{mara}, \ref{romo}) one observes
that $V_{ik}[x]$ containing only one negative value among its
non-diagonal elements cannot be represented as in (\ref{mara}).  Hence
for such a $V_{ik}[x]$ the weak principle holds.  }

\begin{figure}
\includegraphics[width=7cm]{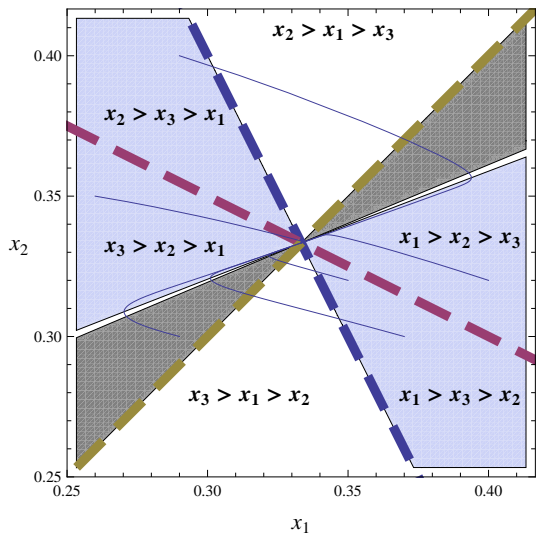}
\caption{(Color online.)
The local structure of (\ref{z2}) around asymptotically stable Nash
equilibrium (\ref{bask}) under conditions (\ref{korkud}). The parameters
in (\ref{nola}) are: $\vartheta=1,\,b_1=2, \,b_2=1.2,\, b_3= 2$. The
Nash equilibrium (\ref{bask}) is asymptotically stable (with real
eigenvalues of the linearized matrix $V_{ik}[p]$) but not globally
stable, another asymptotically stable Nash equilibrium is $x_1=x_2=0$.
The Le Chatelier principle supporting domain is shaded normally (blue). The
domain where the weak principle is violated is shaded dark gray. Domains supporting only the
weak principle are left blank. Three thick dashed lines are $x_2=x_1$,
$x_2=1-2x_1$ and $x_2=(1-x_1)/2$; as indicated, they bound regions with
six different types of ordering. Ordinary lines denote typical
trajectories of (\ref{z2}) moving towards (\ref{bask}) located at the
crossing of three thick dashed lines. 
} 
\label{f1}
\end{figure}

\subsection{Asymptotically, but not globally stable rest-point}

For $n=3$ there are two possibilities for the replicator dynamics with
an asymptotically stable, interior (and generic!) Nash equilibrium $\bp>0$
\cite{zeeman}. {\it (i)} it is globally stable: for all solutions with
$\bx(0)>0$ it holds $\bx(t\to\infty)\to \bp$. {\it (ii)} One of the vertices
[vertex is a probability vector whose one component is $1$] is also
asymptotically stable. This is the only vertex that is stable with
respect to the boundary directions, and no other stable rest-points
exist besides this vertex and $\bp$. The vertex is also evolutionary
stable, and thus the interior rest-point is not evolutionary stable;
otherwise it had to be a unique Nash equilibrium \cite{hofbauer}. We now
turn to discussing the case {\it (ii)}.

Let replicators 1 and 2 be mutualistic, but 1 gains more than 2 from this mutualism:
$a_{12}>a_{21}>0$.  Replicator 3 predates on 1 ($a_{13}<0$, $a_{31}>0$),
but competes with 2: $a_{23},a_{32}<0$ (in a sense 2 saves 1 from 3). Hence the vertex $x_3=1$ is
asymptotically stable; see (\ref{z2}). There can be also an internal
asymptotically stable state if (paradoxically) 3 competes and predates strongly enough.
Within parametrization (\ref{nola}) all the above conditions are satisfied for
\BEA
\vartheta=1, ~~~~
b_1>b_2>1, ~~~~ b_3>\frac{b_1b_2-1}{b_1-b_2}>b_2>1,
\label{korkud}
\EEA
where $b_3>\frac{b_1b_2-1}{b_1-b_2}$ comes from the asymptotic stability 
of the internal rest-point (\ref{bask}); see (\ref{ghost}). Eq.~(\ref{korkud}) describes
a generic situation with two asymptotically stable rest-points. Their
attraction basins are detached by a separatrix that joins two rest
points: $x_1=1$ and $(x_1=0,x_2=\frac{2b_2-2}{b_2+b_3-2})$. 

The internal rest-point (\ref{bask}) is asymptotically stabilized by the
assumed mutualism between 1 and 2. However, with respect to the Le
Chatelier principle this rest-point is fragile: there are {\it local}
perturbations that violate the weak principle \footnote{If this system
of three replicators is kept under weak noise, then
(sooner or later) the separatrix will be crosed, the system will appear
in the attraction basin of the vertex $x_3=1$ and irreversibly settle
there. It is suggestive that this decay of the
interior Nash equilibrium will (more probably) take place through the
region, where the weak Le Chatelier principle is violated. }; see
Fig.~\ref{f1}.

There are three types of perturbations for this situation: {\it (1)}
supporting the Le Chatelier principle; {\it (2)} supporting the weak Le
Chatelier principle only; {\it (3)} violating the weak Le Chatelier
principle, i.e., transiently amplifying perturbation. Domains of the
type {\it (1)} can be divided further into subdomains {\it (1.1)}, where
the trajectory remains confined to that domain for all times, i.e.,
(\ref{major}) holds for all $\epsilon>0$, and subdomains {\it (1.2)},
where the short-time response follows the Le Chatelier principle, but
for intermediate times the trajectory leaves the domain. For the case
shown in Fig.~\ref{f1} all the trajectories end up in
subdomains of the type {\it (1.1)}. 
Circulation of trajectories around the Nash equilibrium is excluded, since the
eigenvalues (\ref{ghost}) are real.

\subsection{Rock-scissors-paper [RSP] game}

Cyclically dominant species are observed in nature and have certain
unexpected features \cite{nz,schuster}. The cyclic dominance was proposed to
be one of the basic mechanisms for maintaining diversity; see
\cite{nz,schuster} and refs. therein.  Their simplest model is the RSP
game, where [recall that $a_{kk}=0$]:
\BEA
\label{rsp}
a_{12}, ~~ -a_{21}, ~~ -a_{13}, ~~
a_{31}, ~~ a_{23}, ~~  -a_{32}
\EEA
are all positive meaning cyclic dominance; see (\ref{tutu1}--\ref{tutu3}). 

\begin{figure}
\vspace{0.2cm}
\includegraphics[width=8.7cm]{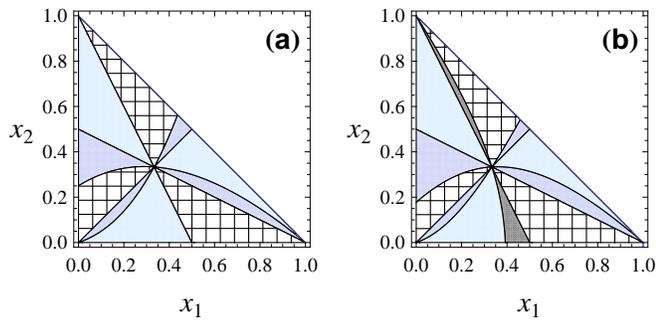}
\vspace{0.3cm}
\caption{(Color online.)
The structure of (\ref{z2}) on the simplex (\ref{ko}) 
with globally stable Nash
equilibrium (\ref{bask}) (bold dot) for $n=3$ and $\vartheta=0.5$; see (\ref{nola}). Fig. 3(a): $b_1=1,
\,b_2=-1,\, b_3= 2$. Fig. 3(b): $b_1=1, \,b_2=-1,\, b_3= 3$. 
The Le Chatelier principle supporting domain is shaded normally (blue).
Light (blue) shaded domains support only the weak principle, where both strongest and weakest components of a perturbation grow.
Meshed domains also support only the weak principle, but now both strongest and weakest components of a perturbation decrease.
Regions violating the weak principle are shaded dark.
 } 
\label{f3}
\end{figure}

The negativity of $a_{21}, \,a_{13},\, a_{32}$ implies that for RSP the
Le Chatelier principle cannot be satisfied for {\it all} local
perturbations; see (\ref{baran}). For the case (\ref{nola}) the RSP
conditions (\ref{rsp}) amount to $b_1>\vartheta>0$, $b_3>\vartheta>0$
and $b_2<-\vartheta<0$.  As the right Fig.~\ref{f3} shows, provided that
$|b_i|$'s are sufficiently different from each other, even the weak Le
Chatelier principle can be violated in the vicinity of the globally
stable Nash equilibrium (\ref{bask}). Thus cyclically dominating species
can amplify small perturbations. The left part of Fig.~\ref{f3} displays
a scenario, where the principle is valid for four particular domains,
while the weak principle holds everywhere. The situation with the zero
sum game $a_{ik}=-a_{ki}$ is very similar to the left Fig.~\ref{f3}.
Here the internal rest point $\bp$ is only neutrally [not asymptotically]
stable, i.e. $\bx(t)$ does not converge to $\bp$. However, 
the time-average converges to $\bp$: $\int_0^T\frac{\d
t}{T}\,\bx(t)\to \bp$ for $T\to \infty$ \cite{hofbauer}. Although the notion
of asymptotic stability does not apply, the Le Chatelier principle stays
well-defined.  We see an interplay between various notions of stability:
RSP-replicators can violate the weak principle around an asymptotically
stable rest-point [see the right Fig.~\ref{f3}], but neutrally stable
zero-sum replicators satisfy the weak principle.

\subsection{Summary}

We now briefly summarize results obtained for three replicators for
perturbations over an asymptotically stable, interior (and unique) Nash
equilibrium. 

\begin{itemize}

\item The Nash equilibrium is always in (or at the boundary of) the
domain, where the Le Chatelier principle holds. 

\item The principle holds locally (i.e., for sufficiently weak
perturbations) under condition (\ref{baran}) that implies mutualism. 

\item Whenever the principle holds locally, all other
perturbations do not destabilize the system. 

\item Violations of the principle (or of its weak form) can show up
already for arbitrary weak perturbations. 

\item The weak principle holds for an evolutionary stable Nash
equilibrium. Even for such an equilibrium, the principle can be violated
locally. 

\item The weak principle (and for certain perturbations the full principle) 
does hold for zer-sum replicators, whose Nash equilibrium is only neutrally
stable.

\end{itemize}

\section{Conclusion}
\label{konkin}

In this paper we applied the Le Chatelier principle to replicator
dynamics of Evolutionary Game Theory (EGT) aiming to understand how
stability features of a Nash equilibrium resemble those of the
thermodynamic equilibrium. Analogies between these two type of
equilibria were noted at several instances \cite{bertin,polish}; e.g.,
they both satisfy the notion of detailed balance (\ref{db}). 

The Le Chatelier principle states that a perturbation of an equilibrium
state starts to diminish already on short times, in contrast to
asymptotic stability, where perturbations are supposed to decay sooner
or later. The Le Chatelier principle had several predecessors in natural
philosophy \cite{phi} and is widely known beyond its original
application domain of quasi-equilibrium thermodynamics
\cite{edu,econo,nature,jorge,gorshkov}.
However, its quantitative formulations is not widely known \footnote{Even within its original domain the formulation
of the Le Chatelier has several delicate (but rather important) points that are normally not recognized in literature; see
\cite{gilmore} for a careful derivation of the principle within quasi-equilibrium thermodynamics.}.  Hence our first
step was to reformulate the Le Chatelier principle for a non-linear
master equation|a common framework for statistical physics and
population dynamics|and relate it to the majorization concept. We were
led to distinguish between the proper Le Chatelier principle
(perturbation induces negative feedback) and its weak vesion
(perturbation does not induce strong positive feedback).

Our results led to refining the notion of Nash equilibrium and
evolutionary stability. The first result is that the validity of the Le
Chatelier principle relates to mutualism. Other notions of stability
(e.g., asymptotic stability) do not allow to indicate in which sense the
mutualism can be more stable than other forms of interactions that also
lead to an asymptotically stable [or even globally stable] interior
rest-point \footnote{The notion of the evolutionary stability can be
related to the mutualism in the whole; see (\ref{golem}. However, the
latter concept does connect in any direct way to mutualistic inter-group
interactions in the sense of (\ref{tutu1}).}. This conclusion points at
a stability-based mechanism for explaining the emergence and maintenance
of mutualistic interactions. 

More generally, we noted that the viewpoint of the Le Chatelier
principle allows to uncover rich scenarios of behaviour for a stable
Nash equilibrium under perturbations.  There are always perturbation
domains, where the Le Chatelier principle is satisfied, but there are
also perturbation domains, where only the weak principle holds.
Moreover, certain non-mutualistic [e.g., cyclically competing]
replicators with a single globally stable Nash equilibrium are capable
of violating even the weak Le Chatelier principle, i.e., they are
endowed with a strong form of positive feedback (weaker forms of
positive feedback are possible already when only the weak principle
holds). This is impossible for perturbations over a thermodynamic
equilibrium, since there it would mean a negative entropy production. 

The existence of positive feedback mechanisms is recognized in ecology
at several instances. It is argued that younger [on the evolution scale]
natural ecosystems have more positive feedback mechanisms than older
ones, where negative feedback dominates \cite{svi}. In macroevolution
dynamics positive feedback is essential for the punctuated equilibrium:
most species persist for ages without showing any significant
morphological change [equilibrium], but they do display relatively
sudden changes via a positive-feedback response after specific
perturbations \cite{punctuated_equilibrium}. Hence before the
punctuation starts, the equilibrium state should be stable, but it
cannot be "equally stable" with respect to all possible perturbations.
Clearly, this combination of various forms of stability (or rather various
forms of feedback) can be quantified via the Le Chatelier principle. 

We focussed on perturbations over an asymptotically stable (interior)
Nash equilibrium. The Le Chatelier principle applies to replicators
satisfying a weaker form of stability: {\it permanence} (or Lagrange
stability) that rules out the extinction of any replicator that was present
in the population initially \cite{svi,hofbauer}. Some important classes
of replicators|e.g., those playing an asymmetric [two-population] or a
zero-sum game|can be only permanent, e.g., for them a Nash equilibrium
is not asymptotically stable (or even not stable, as for hypercyclic
replicators) \cite{hofbauer,sandholm}. One problem with this type of
stability is that a sequence of relatively weak perturbations may drive
replicators towards the simplex boundary, where the extinction risks are
high \cite{justus}. The weak Le Chatelier principle will guarantee the
absence of such a scenario, and its application to this situation is
straightforward, because a permanent system of replicators (\ref{z2})
does have a unique (but generally not stable) interior Nash equilibrium
$\bp>0$, with time-averaged state converging to $\bp$: $\int_0^T\frac{\d
t}{T}\,\bx(t)\to \bp$ for $T\to \infty$ \cite{hofbauer}. An example of
studying permanence via the weak Le Chatelier principle was given above
for a zero-sum game. 

\comment{Another possible extension of the Le Chatelier principle can be
given for more general (e.g., Lotka-Volterra) models that describe the
population dynamics in terms of the group sizes $N_k$
\cite{svi,hofbauer}. Here perturbations will in general not keep the sum
$\sum_{k=1}^n N_k(t)$ constant in time. In this case one can relate the
Le Chatelier principle to the concept of weak super-majorization
\cite{olkin}. }

Here we restricted ourselves with perturbations of the state, i.e., the
probability vector $\bx$.  The Le Chatelier principle should also be
studied for structural perturbation that do not act directly on the
state $\bx$, but influence the structure of the system making the
transition probabilities $W_{i\not=k}[\bx,\lambda_t]$ time-dependent via
a parameter $\lambda_t$; see \cite{ives} for the relevance of such
perturbations in ecology. 

There is a long history of applying thermodynamic stability principles
to bio-ecological systems; see
\cite{trincher,ulan,critics,chakra,karo,rubin_fokht}.  These studies are
definitely thought-provoking, e.g., an observation by Trincher
\cite{trincher} that the qualitative statement of the Le Chatelier
principle is to be violated in the processes of embryogenesis inspired
the present work (cf. with the above discussion on the punctuated
equilibrium). 

However, the basic tool in almost all such applications is the notion of
entropy production, whose features (e.g., positivity) are invoked for
making predictions about the stability of a bio-ecological system. This
approach lacks operationalism, because for the entropy production to be
well defined one should assume that the considered bio-ecological system
is literally thermodynamical, i.e., that concepts developed within
physics (thermodynamic equilibrium, the notions of entropy, temperature,
forces and fluxes) apply. Obviously, an open and non-equilibrium
bio-ecological system produces entropy somewhere and somehow, but one
needs specific conditions for applying concepts developed within
physics; in particular, it has to be verified that the formally
introduced entropy production indeed has a {\it physical} meaning. This
is best examplified by the current literature, e.g.
Refs.~\cite{chakra,karo,rubin_fokht,andrae} that claim to apply the
entropy production concept to the same population dynamic models, {\it
all} employ different definitions of this concept (albeit some of these
definitions become equivalent in the vicinity of the rest-point; see
\cite{chakra,rubin_fokht,andrae}). In contrast, the presented
application of Le Chatelier principle is based on the operational notion
of perturbation decay. Hence, its application need not assume any
would-be-valid thermodynamic reasoning. 

{\bf Acknowledgement}.
This work has been supported by Volkswagenstiftung.

\end{document}